\documentstyle[psfig]{aipproc}

\begin{document}
\title{Compact Binary Mergers and Accretion-Induced Collapse: Event Rates}

\author{Vassiliki Kalogera}
\address{Harvard-Smithsonian Center for Astrophysics \\ Cambridge,
Massachusetts 02138}


\maketitle

\begin{abstract}
 This paper is a brief review of the topic of binary systems as sources of
gravitational-wave emission for both LIGO and LISA. In particular I review
the current estimates of the associated Galactic event rates and their
implications for expected detection rates. I discuss the estimates for (i)
the coalescence of close binaries containing neutron stars or black holes,
(ii) white dwarfs going through accretion-induced collapse into
neutron stars, and (iii) detached but close binaries containing two white
dwarfs. The relevant uncertainties and robustness of the estimates are
addressed along with ways of obtaining conservative upper limits.
 \end{abstract}

\section*{Introduction}

An important factor in the design and development of gravitational wave
observatories is the prospect for detection of astrophysical systems known
or expected to be sources of gravitational radiation. At an initial level,
even qualitative knowledge of the source properties dictates the frequency
range of operation and the desired sensitivity levels of the instruments.
More detailed understanding of signal characteristics, such as waveforms
and polarization, allows the development of optimized data analysis
techniques, followed by early testing and calibration of the system based
on model data. Hence, studies of several different astrophysical sources
of gravitational radiation is an integral part of the collective effort
for the {\em direct} detection of gravitational waves in the near future.

The inspiral of close binary compact objects, neutron stars (NS) or black
holes (BH), driven by gravitational wave emission is considered one of the
major sources for ground-based laser interferometers, such as LIGO, VIRGO,
GEO600, and TAMA300. At present only systems containing two neutron stars
(NS) have been detected, PSR B1913+16 being the prototypical NS--NS system
\cite{HT75}. This binary radio pulsar has provided striking empirical
confirmation of general relativity with the measurement of orbital decay
due to gravitational radiation \cite{TW82}. As this decay proceeds, both
the amplitude and characteristic frequency of the gravitational-wave
signal increase. Although for PSR B1913+16 the frequency will not enter
the LIGO window for another $\sim 3\times 10^8$\,yr, the expectation is
that similar systems in other galaxies, well ahead in their inspiral
phase, should be detectable now. In addition to NS--NS systems, and based
on theories of binary evolution, BH--NS and BH--BH binaries are also
expected to exist in galaxies, although they have not been discovered yet.

Another type of gravitational-wave source is provided by hot, young NS
formed through the collapse of massive stars or the collapse of white
dwarfs driven beyond the Chandrasekhar limit by accretion from a close
binary companion (accretion-induced collapse). A variety of physical
phenomena (e.g., rotational instabilities) can induce a quadrupole moment
in proto-neutron stars and cause them to emit gravitational waves (see
contributions in these proceedings by S.~Hughes and J.~Houser).

An assessment of detection prospects for the various sources requires
estimates of (i) the signal strength, and hence the maximum distance out
to which sources could be detected given the instrument sensitivity, and
(ii)  the source formation rate out to that maximum distance, extrapolated
from Galactic formation rate estimates.  In this paper I review current
estimates of Galactic event rates for the coalescence of binary compact
objects (NS--NS, BH--NS, BH--BH) and the formation of young NS in
accretion-induced collapse of white dwarfs. A critical discussion of the
various uncertain factors involved in these estimates is presented. At the
end of the paper, I review briefly current predictions for close binaries
as gravitational-wave sources expected to be detected by the future space
laser interferometer LISA.

\section*{Coalescing Binaries and LIGO}

Estimates of formation rates for {\em coalescing} binary compact objects
(systems with tight enough orbits that will merge due to gravitational
radiation within a Hubble time) can be predicted theoretically, based on
our current understanding of binary evolution models. For NS--NS systems,
we can also obtain empirical rate estimates based on the observed sample.
In what follows I critically review the current coalescence-rate
estimates, addressing the various uncertainties involved. A discussion of
ways to obtain limits on the NS--NS coalescence rate is also included.

Given the expected strength of the gravitational-wave signal of double
NS coalescence, the maximum distance out to which it could be detected
by the LIGO-II interferometers has been estimated to be $\sim 450$\,kpc
\cite{G99}. A Galactic NS--NS coalescence rate of $\sim
10^{-6}$\,yr$^{-1}$ is then required for a detection rate of $2-3$
events per year. The corresponding estimates for the coalescence of two
10\,M$_{\rm \sun}$\, BH are $\sim 2000$\,kpc and $\sim
10^{-8}$\,yr$^{-1}$ (these distance estimates take into account
cosmological corrections for a flat universe and
$H_0=65$\,km\,s$^{-1}$\,Mpc$^{-1}$ \cite{Fi99}).

\subsection*{Theoretical Estimates}

Theoretical calculations of the formation rate of coalescing binaries
are possible, given a sequence of evolutionary stages followed by
primordial binaries.  Over the years a relatively standard picture has
been formed based on the consideration of NS--NS binaries \cite{H76},
although more recently variations of it have also been
discussed~\cite{B95}. In all versions of their formation path the main
picture remains the same. The initial binary progenitor consists of two
binary members massive enough to eventually collapse into NS or BH. Its
evolution involves multiple phases of stable or unstable mass transfer,
common-envelope evolution, and accretion onto neutron stars, as well as
two supernova explosions.

Such theoretical modeling has been undertaken by various authors by
means of population syntheses. This provides us with {\em ab initio}
predictions of the coalescence rate. The evolution of an ensemble of
primordial binaries with assumed initial properties is followed through
specific evolutionary stages until a coalescing binary is formed. The
changes in the properties of the binaries at the end of each stage are
calculated based on our current understanding of the various processes
involved: wind mass loss from massive hydrogen- and helium-rich stars,
mass and angular-momentum losses during mass transfer phases,
dynamically unstable mass transfer and common-envelope evolution,
effects of highly super-Eddington accretion onto neutron stars, and
supernova explosions with kicks imparted to newborn neutron stars. Given
that several of these phases are not very well understood, the results
of population synthesis are expected to depend on the assumptions made
in the treatment of the various processes. Therefore, exhaustive
parameter studies are required by the nature of the problem.

Recent studies of the formation of binary compact objects and
calculations of coalescence
rates (see \cite{L97}, \cite{F98}, \cite{PZ98}, \cite{BB98})
 have explored the model parameter space and the robustness of the
results at different levels of (in)completeness. Almost all have studied
the sensitivity of the coalescence rate to the average magnitude of the
kicks imparted to newborn neutron stars. The range of predicted Galactic
NS--NS rates from {\em all} these studies obtained by varying the kick
magnitude within reasonable ranges is $ < 10^{-7}~-~5\times
10^{-4}$\,yr$^{-1}$. This large range indicates the importance of
supernovae (two in this case) in the evolution of binaries. Variations
in the assumed mass-ratio distribution for the primordial binaries can
{\em further} change the predicted rate by about a factor of $10$, while
assumptions of the common-envelope phase add another factor of about
$10-100$. Variation in other parameters typically affects the results by
factors of two or less. Results for BH--NS and BH--BH binaries lie in
the ranges $< 10^{-7}~-~10^{-4}$\,yr$^{-1}$ and $<
10^{-7}~-~10^{-5}$\,yr$^{-1}$, respectively when the kick magnitude to
both NS and BH is varied. Other uncertain factors such as the critical
progenitor mass for NS and BH formation lead to variations of the rates
by factors of $10-50$.

It is evident that recent theoretical predictions for the coalescence
rates cover a very wide range of values (typically 3-4 orders of
magnitude). We note, however, that binary properties other than the
coalescence rate, such as orbital sizes, eccentricities, center-of-mass
velocities, are much less sensitive to the various input parameters and
assumptions;  the latter affect more severely the absolute normalization
(birth rate) of the population. Given these results it seems fair to say
that population synthesis calculations have quite limited predictive power
and provide fairly loose constraints on coalescence rates.

\subsection*{Empirical Estimates}

In the case of NS--NS binaries, there is another way to estimate their
coalescence rate, using the properties of the observed coalescing NS--NS
(two systems: PSR B1913+16 and PSR B1534+12) and models of selection
effects in radio pulsar surveys. For each observed object, a scale factor
is calculated based on the fraction of the Galactic volume within which
pulsars with properties identical to those of the observed pulsar could be
detected, in principle, by any of the radio pulsar surveys, given their
detection thresholds. This scale factor is a measure of how many more
pulsars like those detected in the coalescing NS--NS systems exist in our
galaxy. The coalescence rate can then be calculated based on the scale
factors and estimates of detection lifetimes summed up for the observed
systems.  This basic method was first used by Phinney~\cite{P91} and
Narayan et al.~\cite{N91} who estimated the Galactic rate to be $\sim
10^{-6}$\,yr$^{-1}$.

Since then, estimates of the coalescence rate have decreased
significantly primarily because of (i) the increase of the Galactic
volume covered by radio pulsar surveys with no additional coalescing
NS--NS binaries discovered \cite{CL95}, (ii) the increase of the
distance estimate for PSR B1534+12 based on measurements of
post-Newtonian parameters \cite{S98} (iii) changes in the lifetime
estimates for the observed systems \cite{HL96}, \cite{A99}. On the other
hand, in these recent studies a upward correction has been added to
account for the population of pulsars too faint to be detected by the
surveys. The most recently published study~\cite{A99} gives a lower
limit of $2\times 10^{-7}$\,yr$^{-1}$ and a ``best'' estimate of $\sim
6-10\times 10^{-7}$\,yr$^{-1}$ which agrees with other recent estimates
of $2-3\times10^{-6}$\,yr$^{-1}$ \cite{S98}, \cite{E99}. Additional
uncertainties (typically by factors $\lesssim 10$) arise from the
estimates of pulsar ages and distances, the pulsar beaming fraction, the
spatial distribution of NS--NS binaries in the Galaxy, the form of the
faint end of the luminosity function and the small number of objects in
the observed sample.

Despite all these uncertainties the empirical estimates of the NS--NS
coalescence rate appear to span a range smaller than two orders of
magnitude, which is relatively narrow compared to the range covered by the
theoretical estimates.

\subsubsection*{Small-Number Sample}

One important limitation of empirical estimates of the coalescence rates
is that they are derived based on {\em only two} observed NS--NS
systems, under the assumption that the observed sample is representative
of the true population, particularly in terms of their radio luminosity.
Therefore, assessing the effect of small-number statistics on the
results of the above studies is necessary. Assuming that NS--NS pulsars
follow the radio luminosity function of young pulsars and that therefore
their population is dominated in number by low-luminosity pulsars, it
can be shown that the current empirical estimates most probably
underestimate the true coalescence rate.  If a small-number sample is
drawn from a parent population dominated by low-luminosity (hence hard
to detect) objects, it is statistically more probable that the sample
will actually be dominated by objects from the high-luminosity end of
the population. Consequently, the empirical estimates based on such a
sample will tend to overestimate the detection volume for each observed
system, and therefore underestimate the scale factors and the resulting
coalescence rate.

\begin{figure}[b!] 
\centerline{\psfig{file=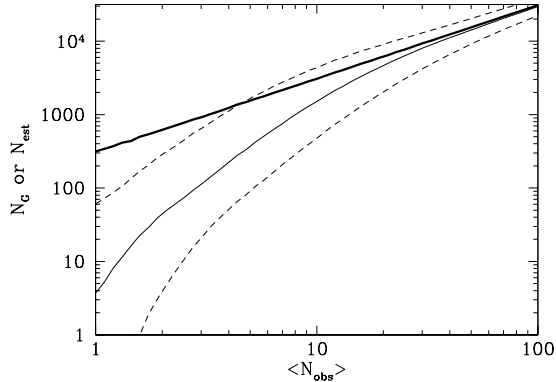,height=2.in,angle=-90.0}}
\vspace{10pt}
\caption{Bias of the empirical estimates of the NS--NS coalescence rate
because of the small-number observed sample. See text for details.}
\label{fig1}
\end{figure}

This effect can be clearly demonstrated with a Monte Carlo experiment
\cite{KN99} using simple models for the pulsar luminosity function and
the survey selection effects. As a first step, the average observed number
of pulsars is calculated given a known ``true'' total number of pulsars in
the Galaxy (thick-solid line in Figure \ref{fig1}). As a second step, a
large number of sets consisting of ``observed'' (simulated) pulsars
drawn from a Poisson distribution of a given mean number ($<N_{\rm obs}>$)
with assigned luminosities according to the assumed luminosity function
are realized using Monte Carlo methods. Based on each of these sets, one
can estimate the total number of pulsars in the Galaxy using empirical
scale factors, as is done for the real observed sample. The many
(simulated) `observed' samples can then be used to obtain the distribution
of the estimated total Galactic numbers ($N_{\rm est}$) of pulsars. The
median and 25\% and 75\% percentiles of this distribution are plotted as a
function of the assumed number of systems in the (fake) `observed' samples
in Figure \ref{fig1} (thin-solid and dashed lines, respectively).

It is evident from Figure \ref{fig1} that, in the case of small-number
observed samples (less than $\sim 10$ objects), it is highly probably that
the estimated total number, and hence the estimated coalescence rate, is
underestimated by a significant factor. For a two-object sample, for
example, the true rate maybe higher by more than a factor of ten. This
correction factor associated with the faint-end of the luminosity function
should be applied to the estimated NS--NS coalescence rate in place of the
factor of $\sim 10$ used so far from a direct extrapolation of the
luminosity function.

\subsubsection*{Limits on Coalescence Rates}

One way to circumvent the uncertainties involved in the estimates of the
NS--NS coalescence rate is to focus on obtaining upper or lower limits
to this rate. Depending on how their value compares to the value of
$\sim 10^{-6}$\,yr$^{-1}$ needed for a few LIGO-II events per year, such
limits can provide us with valuable information about the prospects of
gravitational-wave detection.

Bailes \cite{B96} used the absence of any young pulsars detected in
NS--NS systems and obtained a rough upper limit to the rate of $\sim
10^{-5}$\,yr$^{-1}$, while recently Arzoumanian {\it et al.} \cite{A99}
reexamined this in more detail and claimed a more robust upper limit of
$\sim 10^{-4}$\,yr$^{-1}$.

\begin{figure}[b!] 
\centerline{\psfig{file=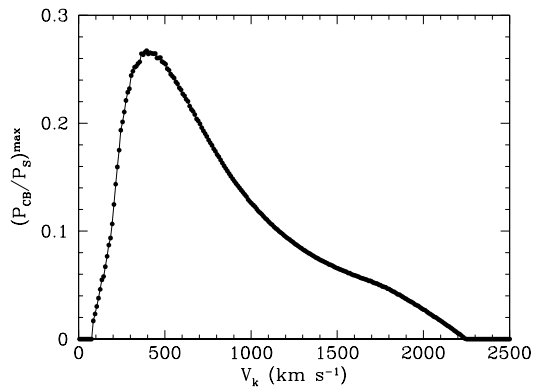,height=2.in,angle=-90.0}}
\vspace{10pt}
\caption{Maximum probability ratio for the formation of coalescing
NS--NS binaries and the disruption of binaries as a function of the
kick magnitude at the second supernova.}
\label{fig2}
\end{figure}

An upper bound to the rate can also be obtained by combining our
theoretical understanding of orbital dynamics (for supernovae with
neutron-star kicks occurring in binaries)  with empirical estimates of the
birth rates of {\em other} types of pulsars related to NS--NS
formation~\cite{KL99}. Binary progenitors of NS--NS systems experience two
supernova explosions when the neutron stars are formed. The second
supernova explosion (forming the neutron star that is {\em not} observed
as a pulsar) provides a unique tool for the study of NS--NS formation,
since the post-supernova evolution of the system is simple, driven only by
gravitational radiation. There are three possible outcomes after the
second supernova: (i) a coalescing binary is formed (CB), (ii) a wide
binary (with a coalescence time longer than the Hubble time) is formed
(WB), or (iii) the binary is disrupted (D) and a single pulsar similar to
the ones seen in NS--NS systems is ejected. Based on supernova orbital
dynamics we can calculate the probability branching ratios for these three
outcomes, $P_{\rm CB}$, $P_{\rm WB}$, and $P_{\rm D}$. For a given kick
magnitude, we can calculate the maximum ratio $(P_{\rm CB}/P_{\rm D})^{\rm
max}$ for the complete range of pre-supernova parameters defined by the
necessary constraint $P_{\rm CB}\neq 0$ (Figure \ref{fig2}). Given that
the two types of systems have a common parent progenitor population, the
ratio of probabilities is equal to the ratio of the birth rates $(BR_{\rm
CB}/BR_{\rm D})$.

We can then use (i) the absolute maximum of the probability ratio
($\simeq 0.26$ from Figure \ref{fig2}) and (ii) an empirical estimate of
the birth rate of single pulsars similar to those seen in NS--NS systems
based on the current observed sample to obtain an upper limit to the
NS--NS coalescence rate. The selection of this small-number sample
involves some subtleties~\cite{KL99}, and the analysis shows $BR_{\rm
CB} \lesssim 1.5\times 10^{-5}$\,yr$^{-1}$~\cite{KL99}. Note that this
number could be increased because of the small-number sample and
luminosity bias affecting this time the empirical estimate of $BR_{\rm
D}$ by a factor of $2-6$.

This is an example of how we can use observed systems other than NS--NS to
improve our understanding of their coalescence rate. A similar calculation
can also be done using the wide NS--NS systems instead of the single
pulsars \cite{KL99}.

\subsection*{Conclusions}

A comparison of the various results on the NS--NS coalescence rate
indicates that theoretical estimates based on modeling of NS--NS formation
have a rather limited predictive power. The range of predicted rates
exceeds 3 orders of magnitude and most importantly includes the
``critical'' value of $10^{-6}$\,yr$^{-1}$ required for a LIGO-II
detection rate of 2-3 events per year. This means that at the two edges of
the range the conclusion swings from no detection to many per month. In
other words no firm conclusions can be drawn from these estimates about
the detection prospects of NS--NS coalescence. Empirical estimates, on the
other hand, derived based on the observed sample appear to be more robust
(estimates are all within a factor smaller than 100). Given those we would
expect a LIGO-II detection rate of a few events per year up to even a few
tens of events per year.

For coalescence rates of BH--NS and BH--BH systems we have to rely solely
on our theoretical understanding of their formation. As in the case of
double NS, the model uncertainties are significant and the ranges extend
to more than 2 orders of magnitude. However, the requirement on the
Galactic rate so that the detection rate is a few events ($2-3$) per year
is less stringent for the BH binaries, only $\sim 10^{-8}$\,yr$^{-1}$.
Therefore, even with the pessimistic estimates for BH--BH coalescence
rates ($\lesssim 10^{-7}$\,yr$^{-1}$), we would expect at least a few or
even up to 10 detections per year, which is quite encouraging. We note
that a very recent examination of dynamical BH--BH formation \cite{PZ99}
in globular clusters leads to detection rates as high as a few per day.

Our expectations for the detectability of BH--NS coalescence could be
improved significantly if we actually detect one or more such systems in
the near future. Current pulsar surveys, such as the Multibeam Parkes
Survey, are considerably more sensitive than previous searches to distant
and faint pulsars in close binaries. This high sensitivity is the combined
result of long integration times, rapid sampling, and the incorporation of
acceleration searches in the data analysis techniques. A candidate NS--NS
system (PSR J1811-1736) has been already discovered \cite{M99} and it
would not be surprising if in the next few years close BH--NS systems are
discovered.

\section*{Young NS in Accretion-Induced Collapse}

Hot, young NS formed in supernovae are susceptible to a range of
instabilities each of which has a different physical origin: e.g.,
convection, bar-mode (dynamical or secular), r-mode instabilities. These
instabilities can induce a quadrupole moment to the proto-NS and cause
it to emit gravitational waves. The strength of the emission varies
significantly with the physical mechanism and its calculation depends on
model assumptions made (for a review see \cite{T96}).  A mechanism which
recently has attracted attention is related to the exponential growth of
rotational mode (r-mode) instabilities in fast spinning ($\lesssim
10-15$\,ms), hot ($>10^9$\,K) proto-NS (see contribution in these
proceedings by B.~Owen, \cite{L98}, and \cite{An99}).  Other mechanisms
involve the centrifugal hang-up of a rapidly spinning collapsing core at
early or late stages of the collapse \cite{T96}.  Such rapidly
rotating proto-NS can be formed either during the collapse of massive
stars, provided that the stellar core is not well coupled to the
envelope and is spinning fast just prior to collapse (see \cite{SP98}),
or perhaps in accretion-induced collapse (AIC) of fast spinning white
dwarfs (WD) in close binaries that accrete at appropriate mass transfer
rates.

Although the details of the the growth of the r-mode instability and the
processes (hydrodynamic or gravitational radiation) for the removal of
the excess angular momentum of rapidly rotating cores are not yet fully
understood, it is interesting to consider these possibilities, primarily
because the signal strength could be high enough for such sources to be
detected by the LIGO-II interferometers out to $\sim 20$\,Mpc for the
r-mode instability \cite{O98} and late centrifugal hang-up at $\sim
20$\,km, and out to $\sim 100$\,Mpc for early centrifugal hang-up at
$\sim 100$\,km \cite{T96}. Adopting the extragalactic extrapolation used
by Phinney \cite{P91}, we find that for a detection rate of a few events
per year Galactic AIC rate of $\sim 10^{-4}$\,yr$^{-1}$ and $\sim
10^{-5}$\,yr$^{-1}$ are required, respectively.

As in the case of binary compact objects, formation of accreting WD that
are expected to go through AIC can be studied via binary population
synthesis techniques and theoretically predicted event rates can then be
obtained. The accuracy and robustness of these results are actually
significantly better than for binary NS and BH because they solely depend
on our understanding of WD formation, which is considerably better than
that of NS and BH formation with birth kicks. The most recent theoretical
study of accreting WD formation \cite{YL98} includes the most up-to-date
picture for the conditions (WD mass and mass transfer rate) necessary for
AIC to occur. The predicted AIC rates for a wide variety of models lie in
the range $8\times 10^{-7} - 8\times 10^{-5}$\,yr$^{-1}$.

Recently an alternative way of addressing the question of AIC event rate
has been explored by \cite{F99}. They use the measured abundances of
neutron-rich nuclei (e.g., $^{62}$Ni, $^{66}$Zn, $^{87}$Rb, $^{88}$Sr)
to set an {\em upper limit} on the AIC rate. This study includes a
detailed investigation of the effect of a number of uncertain factors
(equation of state, neutrino transport, etc.) and the conclusion is that
the upper limit lies in the range $10^{-7} - 10^{-4}$\,yr$^{-1}$. 

Both the actual estimates and the empirical upper limits from the above
studies appear to be in agreement and this gives us confidence that our
understanding of the frequency of AIC events is relatively good.
Comparison with the required Galactic rates for a few events detected
per year indicates that it may be difficult to actually detect
gravitational waves from rapidly rotating proto-NS. However, the upper
end of our estimates is high enough to justify further consideration of
AIC as candidate sources of gravitational radiation.

\section*{Close Binaries and LISA}

The construction of a space laser interferometer, LISA, is also being
planned for the next one or two decades. The frequency range of its
operation is expected to be $\sim 0.1-10^3$\,mHz, clearly different than
that of the ground-based interferometers that are limited by seismic noise
at such low frequencies. As a consequence, LISA will be sensitive to very
different types of gravitational-wave sources and allow us to explore
different regimes of gravity and astrophysical processes.

Although the primary source-target for LISA are supermassive black holes
in the centers of galaxies, LISA will also be sensitive to emission from a
large population of relatively close binaries in our Galaxy \cite{H90}.
For frequencies exceeding $\sim 0.1$\,mHz, this emission is dominated by
close binaries consisting of two white dwarfs. At frequencies up to $\sim
3-4$\,mHz the emission from a large collection of WD--WD binaries is
blended and appears as a continuous background. For higher frequencies up
to $\sim 30$\,mHz, individual sources could be detected, and at even
higher frequencies the extragalactic background becomes detectable but at
a much lower power (a factor of $\sim 10$). 

It is only very recently that WD--WD binaries have been discovered
\cite{M95}, despite the large expected number in the Galaxy.  Their
identification requires challenging optical observations that explain the
small number of objects detected. With such a limited observed sample
estimates of their true number in the Galaxy, and hence of the strength of
the associated GW background, rely on purely theoretical calculations of
their formation (population synthesis). As in the case of AIC models, the
results of these studies are quite reliable since our understanding of the
formation process for WD--WD binaries is better than that of NS and BH.
This is clearly suggested by a comparison of the results from a number of
different recent studies and for different model assumptions \cite{N99},
\cite{WH98}. The formation rate estimates span a relatively narrow range
(factor of 2) from $5\times10^{-3}$\,yr$^{-1}$ to $0.1$\,yr$^{-1}$. The
predicted GR background is largely insensitive to model variations with
the most significant factor being that of mass-ratio distribution in the
primordial binaries.



 





\begin{references}

\bibitem{HT75} Hulse, R.A., and Taylor, J.H., {\it ApJ} {\bf 195}, L51
(1975). 

\bibitem{TW82} Taylor, J.H., and Weisberg, J.M., 
{\it ApJ} {\bf 253}, 908 (1982).

\bibitem{G99} Gustafson, E., Shoemaker, D., Strain, K., and Weiss, R., 
{\it LSC White Paper on Detector Research and Development} (LIGO-Project
document, September 11, 1999).

\bibitem{Fi99} Finn, L.S., {\it private communication} (1999).

\bibitem{H76} van den Heuvel, E.P.J., {\it Structure and Evolution of   
Close Binary Systems}, Dordrecht:  Kluwer Academic Publishers, 1996, pp.
100-100.

\bibitem{B95} Brown, G.E., {\it ApJ} {\bf 440}, 270 (1995).

\bibitem{L97} Lipunov, V.M., {\it et al.}, {\it MNRAS} {\bf 288},
245 (1997).

\bibitem{F98} Fryer, C.L., {\it et al.}, {\it ApJ} {\bf 496}, 333
(1998).

\bibitem{PZ98} Portegies-Zwart, S.Z., and Yungel'son, L.R., 
{\it A\&A} {\bf 332}, 173 (1998).

\bibitem{BB98} Brown, G.E., and Bethe, H., {\it ApJ} {\bf 506}, 780
(1998).

\bibitem{P91} Phinney, E.S., {\it ApJ} {\bf 380}, 17 (1991).

\bibitem{N91} Narayan, R., {\it et al.}, {\it ApJ} {\bf 379}, 17
(1991).

\bibitem{CL95} Curran, S.J., and Lorimer, D.R., {\it MNRAS} {\bf 276}, 347
(1995).

\bibitem{S98} Stairs, I.H., {\it et al.}, {\it ApJ} {\bf 505}, 352
(1998).

\bibitem{HL96} van den Heuvel, E.P.J., and Lorimer, D.R., {\it MNRAS} {\bf
283}, 37 (1996).

\bibitem{A99} Arzoumanian, Z., {\it et al.}, {\it ApJ}, {\bf 520}, 696
(1999).

\bibitem{E99} Evans, T., {\it et al.}, to appear in the proceedings of the
XXXIVth Rencontres de Moriond on "Gravitational Waves and Experimental
Gravity", Les Arcs, France, (1999).

\bibitem{KN99} Kalogera, V., {\it et al.}, in preparation (1999). 

\bibitem{B96} Bailes M., {\it Compact Stars in Binaries}, Dordrecht:
Kluwer Academic Publishers, 1996, pp. 213-223

\bibitem{KL99} Kalogera, V., and Lorimer, D.R., {\it ApJ} {\bf 530}, in
press (2000) [astro-ph/9907426].

\bibitem{PZ99} Portegies-Zwart, S.F., and McMillan, S.L.W., {\it ApJ
Letters}, submitted [astro-ph/9910061].

\bibitem{M99} Manchester, R.N., {\it et al.}, to appear in {\it Pulsar
Astronomy - 2000 and Beyond}, 1999. 

\bibitem{T96} Thorne, K.S., {\it Compact Stars in Binaries}, Dordrecht:
Kluwer Academic Publishers, 1996, pp. 153-183.

\bibitem{L98} Lindblom, L., {\it et al.}, {\it PRL} {\bf 80}, 4843 (1998). 

\bibitem{An99} Andersson, N., {\it et al.}, {\it ApJ} {\bf 510}, 846 
(1999). 

\bibitem{SP98} Spruit, H.C., and Phinney, E.S., {\it Nature} {\bf 393},
139 (1998).

\bibitem{O98} Owen, B., {\it et al.}, {\it PRD} {\bf 58}, 084020(1998). 

\bibitem{YL98} Yungel'son, L., and Livio, M., {\it ApJ} {\bf 497}, 168
(1998). 

\bibitem{F99} Fryer, C.L., {\it et al.}, {\it ApJ} {\bf 516}, 892
(1999). 

\bibitem{H90} Hils, D., {\it et al.}, {\it ApJ} {\bf 360}, 75 (1990). 

\bibitem{M95} Marsh, T.R., {\it et al.}, {\it ApJ} {\bf 275}, 828 (1995). 

\bibitem{N99} Nelemans, G., {\it et al.}, to appear in the proceedings
of the XXXIVth Rencontres de Moriond on {\it Gravitational Waves and
Experimental Gravity}, Les Arcs, France, (1999).

\bibitem{WH98} Webbink, R.F., and Han, Z., {\it Laser
Interferometer Space Antenna}, New York: American Institute of
Physics, 1999, pp.61--67. 

\end{references}
\end{document}